\documentclass[%
 reprint,
superscriptaddress,
 amsmath,amssymb,
 aps,
]{revtex4-2}

\usepackage{graphicx}% Include figure files
\usepackage{dcolumn}% Align table columns on decimal point
\usepackage{bm}% bold math
\usepackage{ragged2e}
\usepackage{subcaption}
\usepackage{amsmath}
\usepackage{xspace}
\usepackage{soul}
\usepackage{amsmath,bm}
\usepackage[utf8]{inputenc}
\usepackage{array}
\usepackage[mathscr]{euscript}

\usepackage[hidelinks]{hyperref}
\hypersetup{
  colorlinks   = true, %Colours links instead of ugly boxes
  urlcolor     = blue, %Colour for external hyperlinks
  linkcolor    = blue, %Colour of internal links
  citecolor   = red %Colour of citations
}
\begin{document}

\preprint{APS/123-QED}

\title{Catalytic Crosstalk: Cooperative Enzyme Dynamics in Artificial Crowded Environments}

% Force line breaks with \\
%\thanks{Author to whom correspondence should be addressed: k.dey@iitgn.ac.in}%

\author{Rik Chakraborty}
%\email{chakraborty_rik@iitgn.ac.in}
\affiliation{Laboratory of Soft and Living Materials, Department of Physics, Indian Institute of Technology Gandhinagar, Gandhinagar, Gujarat 382055, India}

\author{Manisha Jhajhria}
\affiliation{Department of Physics, Indian Institute of Science Education and Research Bhopal, Bhopal, Madhya Pradesh 462066, India}

\author{Arnab Maiti}
\affiliation{Laboratory of Soft and Living Materials, Department of Physics, Indian Institute of Technology Gandhinagar, Gandhinagar, Gujarat 382055, India}
\author{Nividha}
\affiliation{Laboratory of Soft and Living Materials, Department of Physics, Indian Institute of Technology Gandhinagar, Gandhinagar, Gujarat 382055, India}
%\author{Sirsha Ganguly}
%\affiliation{Department of Chemical Engineering, Jadavpur University, Kolkata, West Bengal 700032, India}

\author{Priyanka}
\affiliation{Laboratory of Soft and Living Materials, Department of Physics, Indian Institute of Technology Gandhinagar, Gandhinagar, Gujarat 382055, India}

\author{Snigdha Thakur}
\affiliation{Department of Physics, Indian Institute of Science Education and Research Bhopal, Bhopal, Madhya Pradesh 462066, India}

%\author{Uddipta Ghosh}
%\affiliation{Department of Mechanical Engineering, Indian Institute of Technology Gandhinagar, Palaj, Gujarat 382055, India}

\author{Krishna Kanti Dey}
\email{Author to whom correspondence should be addressed:k.dey@iitgn.ac.in}
\affiliation{Laboratory of Soft and Living Materials, Department of Physics, Indian Institute of Technology Gandhinagar, Gandhinagar, Gujarat 382055, India}
\date{\today}% It is always \today, today,

\newcommand{\tend}[1]{\hbox{\oalign{$\bm{#1}$\crcr\hidewidth$\scriptscriptstyle\bm{\sim}$\hidewidth}}}
%tensor 4:
\newcommand{\tenq}[1]{\hbox{\oalign{$\bm{#1}$\crcr\hidewidth$\scriptscriptstyle\bm{\approx}$\hidewidth}}}
             
%-------ABSTRACT---------         
\begin{abstract}

In cellular environments, enzymes operate under densely crowded conditions that often hinder catalytic efficiency by limiting substrate diffusion and essential conformational dynamics. While reports suggest that crowding can often lead to inhibition of enzyme’s catalytic activity, persistent efficiency of cellular biochemistry hints at underlying cooperative mechanisms among these molecules. Here, we experimentally demonstrate catalytic crosstalk between two enzymes - catalase and urease - in artificially crowded environments. Our results reveal that when co-localized in dense media, these enzymes mutually enhance each other’s catalytic activity and dynamic behavior. This cooperative interaction leads to a net increase in reaction rates and mobility, suggesting an emergent many-body effect in enzyme assemblies. Modeling enzymes as dimeric active particles, we propose a minimal simulation framework that qualitatively captures the observed synergy. Our findings show that inter-enzyme cooperation can counteract the detrimental effects of crowding, offering insights into how enzymatic efficiency is sustained in complex biological milieu.
\end{abstract}

\maketitle
%-------Introduction---------
\section{\label{sec:level1}INTRODUCTION}\protect\
Enzymes in living cells operate in densely crowded and structurally heterogeneous environments, where macromolecules, cytoskeletal filaments, and organelles collectively restrict molecular diffusion and conformational dynamics.~\cite{1,2,3,4,5,6,7,8,9,10,11,12} Such crowding can significantly hinder catalytic turnover by limiting substrate access to active sites or by dampening the structural fluctuations required for enzymatic function.~\cite{13,14,15,16,17} While some studies have reported negligible or even positive effects of macromolecular crowding on certain enzymes’ catalytic efficiency, for many others, crowding is associated with a marked decline in the rate of substrate turnover.~\cite{18,19,20,21,22,23} This raises a fundamental question: how do enzymes maintain functional performance in such physically restrictive intracellular conditions?

One possibility is that enzymes do not act in isolation but rather interact cooperatively within larger assemblies or networks. In addition, intracellular environments are not merely crowded but also actively driven: motor proteins such as kinesin, dynein, and myosin operate ceaselessly, transporting cargo and remodeling cytoskeletal structures.~\cite{24,25,26} These active processes generate persistent, non-thermal mechanical fluctuations that differ fundamentally from equilibrium thermal noise.~\cite{27,28,29,30,31,32,33,34,35} Theoretical studies have proposed that such active fluctuations - or mechanical coupling between enzymes - could help maintain or even enhance enzymatic activity.~\cite{36,37} However, direct experimental evidence of cooperative effects among enzymes under crowding remains to be reported.

In this work, we investigate the catalytic and dynamic behavior of two enzymes - catalase and urease - in artificially crowded environments.~\cite{38,39} While in aqueous solution the enzymes act independently, under crowded conditions they exhibit catalytic crosstalk: the presence of one enhances the catalytic activity and fluctuation dynamics of the other. This mutual reinforcement is not due to classical substrate channeling but likely arise from mechanical or hydrodynamic interactions, leading to improved enzymatic performance. This cooperative interaction also manifests in the diffusion of nearby passive tracer particles. In aqueous environments, the diffusion enhancement from both enzymes is approximately additive. However, under crowded conditions, the enhancement significantly exceeds the sum of individual contributions, indicating a cooperative amplification of local dynamics. This non-additive effect suggest a mechanism by which enzyme activity can enhance molecular transport, potentially supplementing motor driven processes over short length scales in crowded environments. These results point to a general principle by which enzyme assemblies sustain molecular mobility and reaction efficiency under crowding.

To rationalize these findings, we develop a minimal model treating enzymes as dimeric active particles generating non-equilibrium fluctuations. The model captures the observed synergy in catalytic rates and tracer mobility. Together, our results provide direct evidence that enzyme-cooperation can overcome crowding-induced limitations, offering insights into how dynamic crosstalk sustain reaction efficiency and transport in living cells.

\section{\label{sec:level2}RESULTS AND DISCUSSIONS}
\begin{figure}[h]
    \centering
    \includegraphics[width=0.52\textwidth]{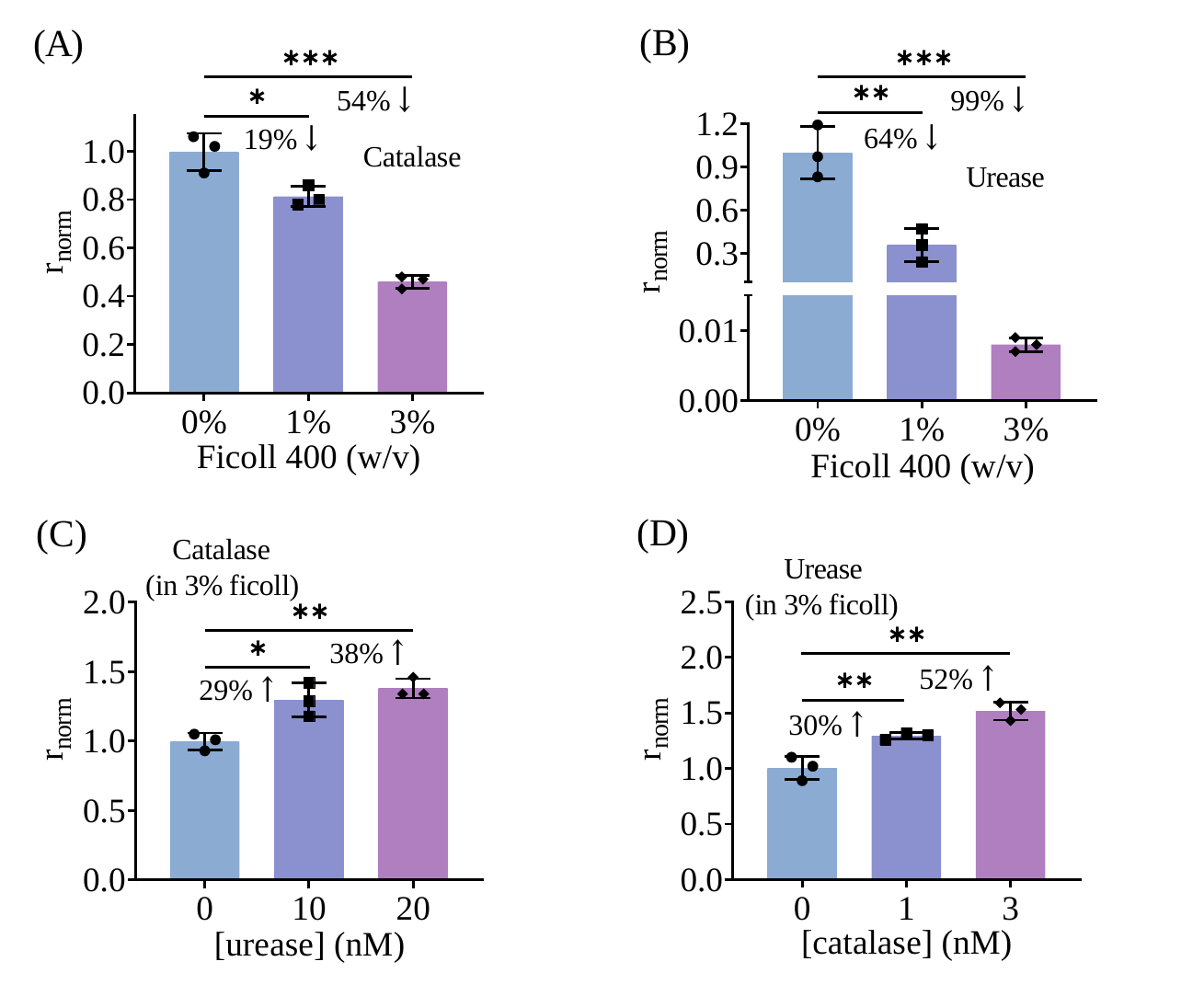}
        \caption{\justifying{Normalized initial reaction rates of (A) catalase and (B) urease as a function of Ficoll 400 concentration (1\% and 3\%), measured over the first 20 s using UV - Vis spectroscopy. Rates are normalized to the value in DI water. The enzyme and substrate concentrations are: [Catalase] = 1 nM, [H$_2$$O_2$] = 10 mM, [Urease]= 10 nM, [Urea] = 100 mM. (C) Catalase activity ([Catalase] = 1 nM, [H$_2$$O_2$] = 10 mM) as a function of urease activity ([Urease]= 10, 20 nM, [Urea] = 100 mM) in 3\% Ficoll (no enhancement in aqueous conditions; see SM). (D) Urease activity ([Urease]= 10 nM, [Urea] = 100 mM) as a function of catalase activity ([Catalase] = 1, 3 nM, [H$_2$$O_2$] = 10 mM) under the same conditions. Error bars denote standard deviation from three independent measurements. Statistical significance: *$p<0.05$, **$p<0.01$, ***$p<0.001$.}}
    \label{fig:fig1}
\end{figure}

\protect\
\textit{Single enzyme activity under crowded conditions}: We measured the catalytic activity of catalase and urease in the presence of the macromolecular crowder Ficoll 400 using UV-Vis spectroscopy (see Supplemental Material, SM). The reaction rate of catalase decreased by 19\% and 54\%, while that of urease declined by 64\% and 99\% at 1\% and 3\% Ficoll, respectively, relative to DI water (Fig.~\ref{fig:fig1}A, B). All rates were normalized to the value in the absence of crowder. Similar trends were observed using Dextran 40, indicating that the suppression was a general feature of crowded environments (Fig. S4 in SM). The reduction in activity might arise from multiple factors, including increased viscosity, altered enzyme conformational dynamics, and restricted substrate diffusion. To isolate the role of viscosity, we performed control experiments with glycerol. Catalase activity decreased by only 18\% in 10\% glycerol, compared to 54\% in 3\% Ficoll, despite comparable viscosities (Fig. S5 in SM), indicating that viscosity alone cannot account for the observed suppression.

 To assess structural effects, we carried out Circular Dichroism (CD) and fluorescence measurements. CD spectra in the far-UV region showed that both enzymes retained their characteristic secondary structures across Ficoll concentrations, with minimal changes in ellipticity at 208 and 222 nm. Fluorescence measurements revealed moderate reductions in intensity ($\sim$23\% for catalase and $\sim$13\% for urease), suggesting only minor  perturbations in the local environment of aromatic residues. Together, these results indicated that the overall structural integrity of the enzymes was largely preserved under crowding (Figs. S6 and S7 in SM). These observations suggested that the dominant effect of crowding was the restriction of enzyme and substrate mobility. Given the comparable sizes of Ficoll and the enzymes, increasing crowder concentration reduced the frequency of productive enzyme–substrate encounters, leading to suppressed reaction rates. However, the differing magnitude of suppression between catalase and urease likely reflected differences in their molecular properties and interactions with the crowded environment.

\textit{Dual-enzyme activity under crowded conditions}: The observed suppression motivated us to explore whether cooperative interactions could offset diffusion-limited constraints. Previous studies showed that enzyme catalysis generated non-thermal mechanical fluctuations that enhanced passive diffusion.~\cite{33,34,35} We hypothesized that such fluctuations could facilitate substrate transport between enzymes,  enabling cooperative enhancement under crowded conditions. To test this, we introduced urease activity into a catalase reaction system in the presence of 3\% Ficoll. Catalase reaction rates increased significantly with increasing urease activity (Fig.~\ref{fig:fig1}C), whereas no enhancement was observed in aqueous conditions. A reciprocal effect was observed for urease in the presence of catalase under crowding (Fig.~\ref{fig:fig1}D), but not in dilute solutions (Fig. S8 in SM). Control experiments under buffered and unbuffered conditions confirmed that this catalytic crosstalk was not driven by pH changes (Fig. S9 in SM).

These results demonstrated that enzyme-generated active fluctuations could mediate non-specific mechanical crosstalk, restoring enzymatic function in crowded environments. Such cooperative interactions were consistent with theoretical predictions of enzyme synchronization and enhanced catalytic rates.~\cite{36,37} We further expected that this cooperative activity would amplify local dynamic fluctuations and enhance transport. As illustrated schematically (Fig. S14 in SM), enzyme-generated fluctuations are more effectively transmitted in crowded environments, suggesting that multi-enzyme systems can produce enhanced, non-additive transport compared to individual enzymes.

\begin{figure}[h] 
    \centering
    \includegraphics[width=0.52\textwidth]{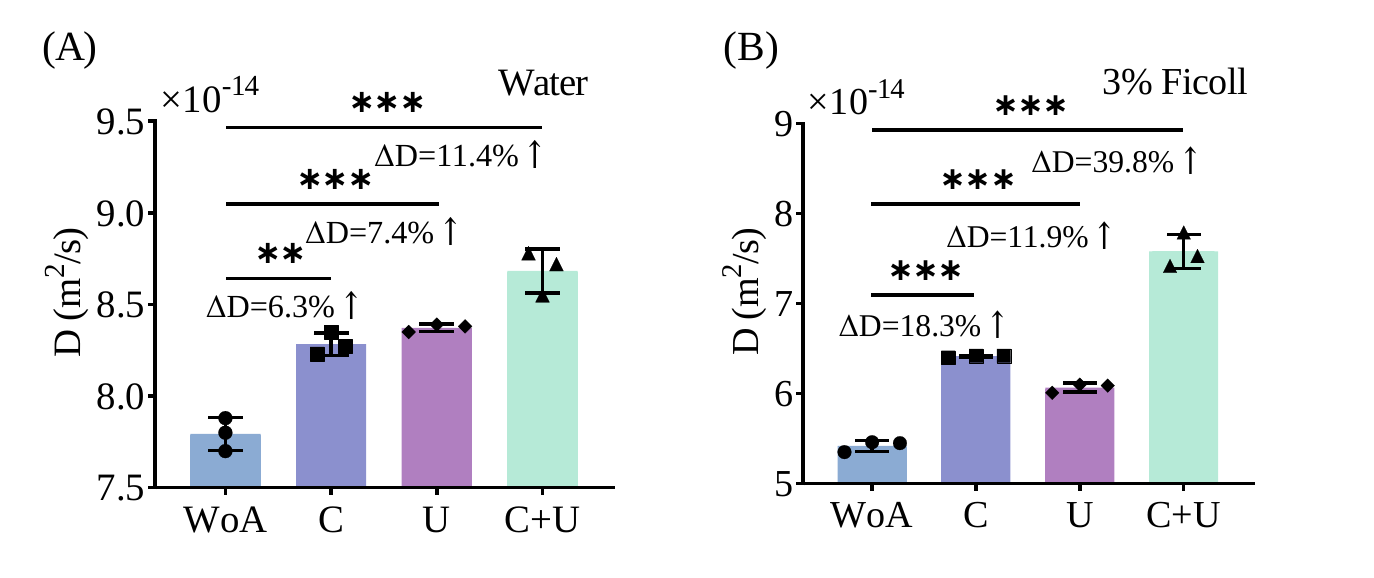}
    \caption{\justifying{Diffusion coefficient of 3 $\mu$m polystyrene tracer particles in the presence of enzymatic activity under (A) aqueous and (B) crowded (3\% Ficoll) conditions. Bars correspond to catalase (C), urease (U), and combined activity (C+U).The enzyme and substrate concentrations are: [Catalase] = 1 nM, [H$_2$$O_2$] = 10 mM, [Urease]= 10 nM, [Urea] = 100 mM. Error bars denote standard deviation from independent experiments (n=3). Statistical significance: **$p<0.01$, ***$p<0.001$.}}
    \label{fig:fig2}
\end{figure}

\textit{Tracer diffusion measurements}: To test the above hypothesis, we measured the diffusion of 3 $\mu$m polystyrene tracer particles in dilute and crowded environments in the presence of enzymatic activity (see SM for details). Previous studies showed that catalase and urease individually enhanced tracer diffusion.~\cite{34} In aqueous solution, catalase and urease increased tracer diffusion by $\sim$6\% and 7\%, respectively, while their combined activity yielded $\sim$12\%, consistent with additivity (Fig.~\ref{fig:fig2}A). In contrast, under crowded conditions (3\% Ficoll), individual enhancements of 18\% (catalase) and 12\% (urease) was increased to $\sim$40\% under simultaneous activity, significantly exceeding the additive expectation (Fig.~\ref{fig:fig2}B).

To quantify this effect, we evaluated $\Delta =\Delta D_{C+U} - (\Delta D_C + \Delta D_U)$, where C and U denoted catalase and urease, and $\Delta D$ was the diffusion enhancement. Under crowded conditions, $\Delta$ was positive and statistically significant, whereas in aqueous solution it was indistinguishable from zero (see SM). Similar non-additive enhancement was observed at higher crowding (5\% Ficoll), where tracer diffusion increased by $\sim$51\% under simultaneous activity (Fig. S13 in SM). These results demonstrated that cooperative enzyme activity under crowding could lead to emergent, non-additive enhancement of tracer mobility, reflecting amplified local fluctuations generated by enzyme crosstalk.

\section{\label{sec:level3}NUMERICAL SIMULATIONS}
\begin{figure}[ht!]
    \includegraphics[width=0.9\linewidth]{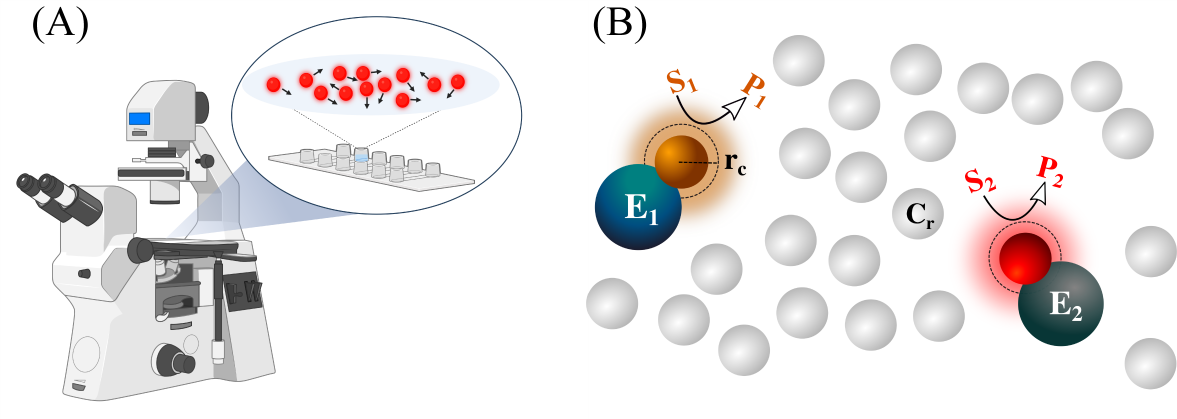}    \caption{\justifying{Schematic of (A) the experimental setup for tracer diffusion measurements and (B) the simulation model. Two dimeric enzyme species, $E_1$ and $E_2$, catalyze reactions at their respective catalytic sites within a reaction zone $r<r_c$. Passive crowder particles ($C_r$) are shown in white.}}
    \label{fig:fig3}
\end{figure}

\protect\
To complement the experiments, we performed numerical simulations of catalytically active enzymes in crowded environments (see SM for details). Following previous studies,~\cite{40,41,42} enzymes were modeled as dimeric active particles composed of catalytic and non-catalytic spherical monomers of diameters $\sigma_{EC}$ and $\sigma_{EN}$, respectively, connected by a harmonic spring potential (Fig.~\ref{fig:fig3}). The non-catalytic monomer was chosen larger to enhance activity.~\cite{43,44} Catalytic sites irreversibly converted substrates into products when a substrate entered a reaction zone ($r<r_c$) around the catalytic monomer ($E_1: S_1 \to P_1$ and $E_2: S_2 \to P_2$). To mimic the quasi-two-dimensional experimental setup (Fig.~\ref{fig:fig3}A), the system was confined within a slab geometry bounded by parallel walls in the z-direction using a harmonic potential,~\cite{41} effectively restricting motion to the xy-plane. Simulations included active enzymes and passive crowders of diameter $\sigma_{C_r}$, (Fig.~\ref{fig:fig3}B) with periodic boundary conditions in the lateral directions. Hydrodynamic interactions were incorporated using a hybrid molecular dynamics–multiparticle collision dynamics (MD–MPCD) scheme. Activity was resulted from self-diffusiophoretic interactions generated by asymmetric product distributions around the enzyme, leading to non-equilibrium fluctuations and propulsion. All particles interacted via short-range repulsive potentials to capture excluded-volume effects. This minimal framework captureed the essential features of enzyme-generated active fluctuations in crowded environments while remaining computationally tractable.

\begin{figure}[h]
    \centering
    \includegraphics[width=0.51\textwidth]{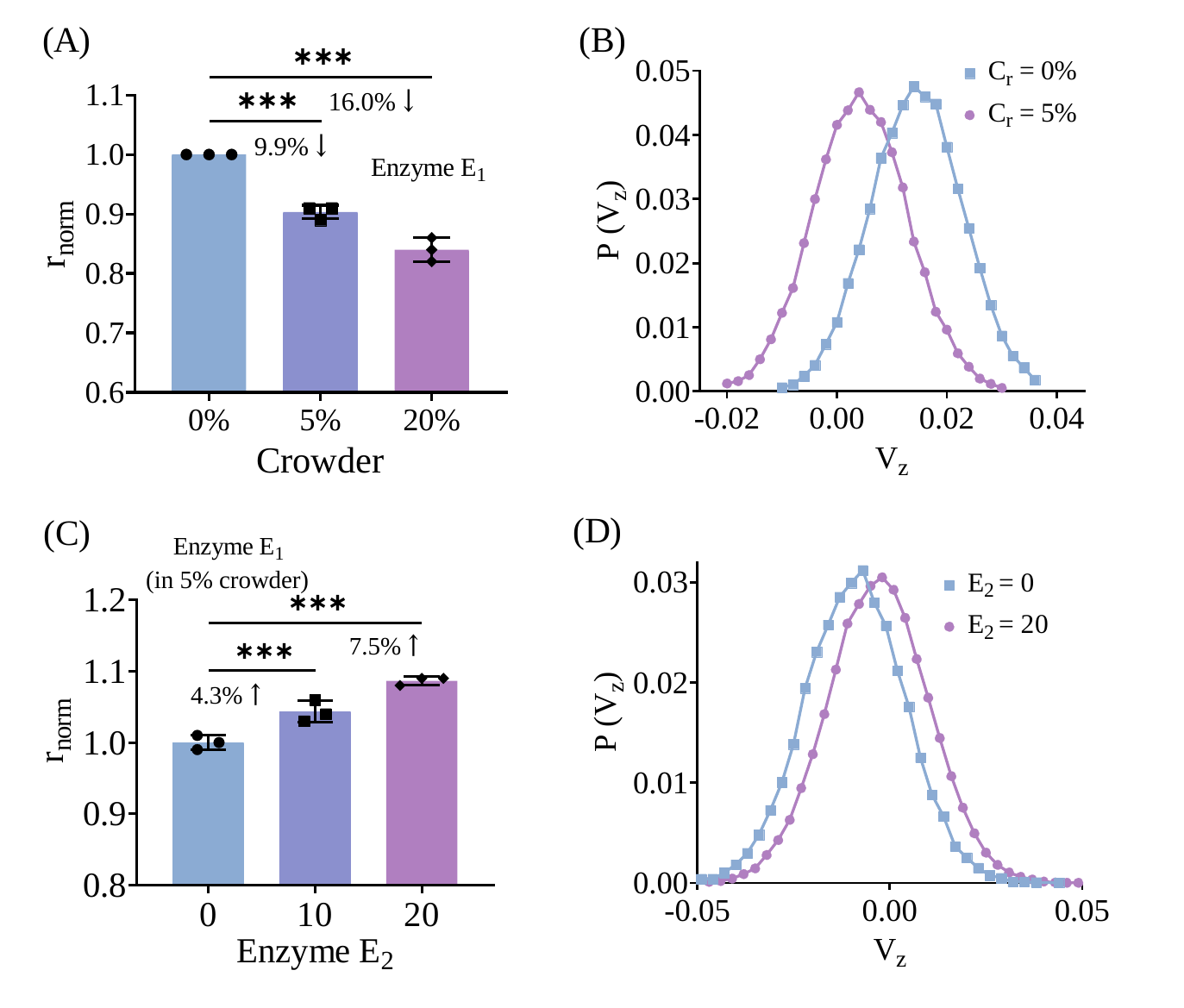}
    \caption{\justifying{(A) Normalized reaction rate of enzyme $E_1$ as a function of crowder concentration. (B) Distribution of instantaneous velocity of $E_1$ along its long axis with and without crowders. (C) Reaction rate of $E_1$ as a function of $E_2$ concentration under crowded conditions (5\% crowder). (D) Distribution of $E_1$ velocity with and without $E_2$ activity under the same conditions. Error bars denote standard deviation from at least 50 independent simulation runs. ***$p<0.001$.}}
    \label{fig:fig4}
\end{figure}

We first examined the effect of crowding on single-enzyme dynamics. Increasing crowder concentration led to a systematic reduction in enzymatic activity (Fig.~\ref{fig:fig4}A), accompanied by a decrease in enzyme mobility (Fig.~\ref{fig:fig4}B). The velocity distribution shifted toward lower values in the presence of crowders, consistent with restricted diffusion of both enzymes and substrates. Although the magnitude of suppression was smaller than in experiments, the simulations reproduced the same qualitative trend, indicating that the crowding limited reaction rates primarily through reduced mobility and encounter frequency.

We next investigated cooperative effects between enzymes by introducing activity of a second species. The reaction rate of $E_1$ increased with increasing activity of $E_2$ (Fig.~\ref{fig:fig4}C), demonstrating catalytic crosstalk. While a modest enhancement was observed in dilute conditions, the effect became significantly stronger under crowding, indicating that cooperative interactions were amplified in restricted environments. This enhancement was accompanied by increased enzyme motility (Fig.~\ref{fig:fig4}D), with velocity distributions shifting toward higher values in the presence of the second active enzyme. These results suggested that activity-induced fluctuations generated by one enzyme could enhance substrate transport and local dynamics experienced by another.

\begin{figure}[t]
    \centering
    \includegraphics[width=0.52\textwidth]{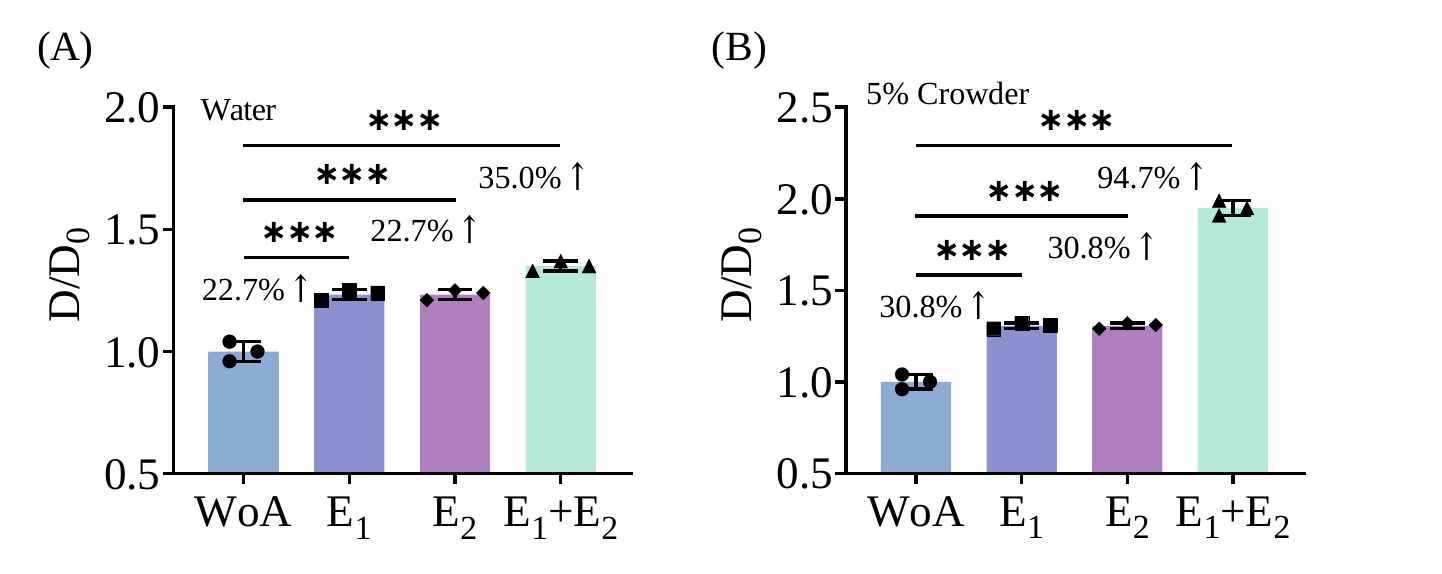}
    \caption{\justifying{(A) Normalized diffusion of passive tracers in the presence of $E_1$, $E_2$, and both enzymes under dilute conditions. The combined effect is approximately additive. (B) Corresponding results under 5\% crowding, where simultaneous activity produces a significantly larger, non-additive enhancement. Error bars denote the standard deviation computed from three averaged MSD measurements (n=3), each obtained from multiple (at least 10) simulation runs. ***$p<0.001$.}}
    \label{fig:fig5}
\end{figure}

To probe the consequences of such crosstalk on transport, we examined tracer diffusion in the presence of enzymatic activity. In dilute conditions, tracer diffusion was enhanced by $\sim$22.7\% in the presence of either $E_1$ or $E_2$ alone and was increased to $\sim$35.0\% when both enzymes were active simultaneously (Fig.~\ref{fig:fig5}A), consistent with an approximately additive response. In contrast, under crowded conditions (5\% crowder), individual enzyme activity enhanced tracer diffusion by $\sim$30.8\%, while simultaneous activity produced a much larger enhancement of $\sim$94.7\% (Fig.~\ref{fig:fig5}B). This strong deviation from additivity demonstrated the emergence of cooperative, non-linear transport behavior in multienzyme systems. 

While the magnitude of diffusion enhancement in simulations exceeded that observed experimentally, the qualitative agreement in the emergence of non-additive behavior indicated that the underlying mechanism was robust. The discrepancies likely arose from the simplified representation of enzymes as rigid dimers and the idealized interaction potentials used in the model. Importantly, tracer motion remained diffusive under all conditions, consistent with experimental observations (see SM, Fig. S16). A key simulation benefit was selectively disabling hydrodynamic interactions via fluid flow decorrelation, without altering enzyme chemical activity. We found that tracer diffusion enhancement persisted even when hydrodynamic interactions were suppressed, indicating that hydrodynamics was not strictly required for mutual reinforcement (see SM, Fig. S18). Overall, these simulations demonstrated that enzyme-generated active fluctuations could mediate non-specific mechanical crosstalk, leading to cooperative enhancement of both catalytic activity and transport under crowded conditions. The results supported a general mechanism by which multienzyme systems could sustain and amplify transport processes in physically constrained environments.

\section{\label{sec:level4}CONCLUSIONS}\protect\
Through complementary experimental and numerical approaches, we demonstrated that non-specific mechanical crosstalk between enzymes in crowded colloidal environments could enhance substrate catalysis for both enzymes and lead to a non-trivial, cooperative increase in passive tracer diffusion. The observed recovery of enzyme reaction rates via cross-enzyme activity, together with the non-additive enhancement of tracer mobility, revealed a previously underexplored mode of collective enzymatic behaviour with functional implications for intracellular transport. In the cytosolic environment, where macromolecular crowding significantly impairs diffusion, both enzyme mobility and substrate accessibility are reduced. While cells rely on energy-consuming motor proteins for long-range transport, local molecular movement over short distances remain largely diffusion-dependent. Our findings suggest that enzyme-generated active fluctuations - especially when reinforced through catalytic crosstalk in multi-enzyme assemblies - can serve as an energy-efficient, passive mechanism to sustain and regulate short-range molecular transport in crowded cellular environments.

\begin{acknowledgments}
\vspace{5mm}
KKD thanks Anusandhan National Research Foundation (ANRF), India (Grants No. CRG/2023/007588 and No. CRD/2024/000753, the latter under the ASEAN- India Collaborative Research and Development Scheme), Ministry of Education, Government of India (Grant No. MoE-STARS/STARS-2/2023-0620), Gujarat State Biotechnology Mission (Grant No. GSBTM/RSS/E-FILE/30/2024/0021/04952485), and IIT Gandhinagar for financial support. ST acknowledges the financial support from ANRF, India (Grant No. CRG/2022/003778).  ST and MJ acknowledge the HPC facility at IISER Bhopal, India, and National Supercomputing Mission (NSM) facility PARAM Shivay at IIT (BHU) Varanasi for allocating the necessary computational resources for conducting the research reported in this paper. RC, MJ, AM and Nividha thank the Ministry of Education, Government of India, and Council of Scientific \& Industrial Research (CSIR) for research fellowships.
\end{acknowledgments}

\bibliography{Manuscript}
\end{document}